\input harvmac

\def\quad{{\ \ }}

\let\includefigures=\iftrue
\newfam\black
\includefigures
\input epsf
\def\figin{\epsfcheck\figin}\def\figins{\epsfcheck\figins}
\def\epsfcheck{\ifx\epsfbox\UnDeFiNeD
\message{(NO epsf.tex, FIGURES WILL BE IGNORED)}
\gdef\figin##1{\vskip2in}\gdef\figins##1{\hskip.5in}
\else\message{(FIGURES WILL BE INCLUDED)}%
\gdef\figin##1{##1}\gdef\figins##1{##1}\fi}
\def\DefWarn#1{}
\def\figinsert{\goodbreak\midinsert}
\def\ifig#1#2#3{\DefWarn#1\xdef#1{fig.~\the\figno}
\writedef{#1\leftbracket fig.\noexpand~\the\figno}%
\figinsert\figin{\centerline{#3}}\medskip\centerline{\vbox{\baselineskip12pt
\advance\hsize by -1truein\noindent\footnotefont{\bf Fig.~\the\figno:}
#2}}
\bigskip\endinsert\global\advance\figno by1}
\else
\def\ifig#1#2#3{\xdef#1{fig.~\the\figno}
\writedef{#1\leftbracket fig.\noexpand~\the\figno}%
#2}}
\global\advance\figno by1}
\fi


\def\sym{ \> {\vcenter {\vbox
{\hrule height.6pt
\hbox {\vrule width.6pt height5pt
\kern5pt
\vrule width.6pt height5pt
\kern5pt
\vrule width.6pt height5pt}
\hrule height.6pt}
}
} \>
}
\def\fund{ \> {\vcenter {\vbox
{\hrule height.6pt
\hbox {\vrule width.6pt height5pt
\kern5pt
\vrule width.6pt height5pt }
\hrule height.6pt}
}
} \>
}
\def\anti{ \> {\vcenter {\vbox
{\hrule height.6pt
\hbox {\vrule width.6pt height5pt
\kern5pt
\vrule width.6pt height5pt }
\hrule height.6pt
\hbox {\vrule width.6pt height5pt
\kern5pt
\vrule width.6pt height5pt }
\hrule height.6pt}
}
} \>
}


\lref\CallanKZ{
  C.~G.~Callan and J.~M.~Maldacena,
  ``Brane dynamics from the Born-Infeld action,''
  Nucl.\ Phys.\  B {\bf 513}, 198 (1998)
  [arXiv:hep-th/9708147].
}

\lref\GibbonsXZ{
  G.~W.~Gibbons,
  ``Born-Infeld particles and Dirichlet p-branes,''
  Nucl.\ Phys.\  B {\bf 514}, 603 (1998)
  [arXiv:hep-th/9709027].
}
\lref\HoweUE{
  P.~S.~Howe, N.~D.~Lambert and P.~C.~West,
  ``The self-dual string soliton,''
  Nucl.\ Phys.\  B {\bf 515}, 203 (1998)
  [arXiv:hep-th/9709014].
}

\lref\HoweET{
  P.~S.~Howe, N.~D.~Lambert and P.~C.~West,
  ``The threebrane soliton of the M-fivebrane,''
  Phys.\ Lett.\  B {\bf 419}, 79 (1998)
  [arXiv:hep-th/9710033].
}
\lref\BergshoeffBH{
  E.~Bergshoeff, J.~Gomis and P.~K.~Townsend,
  ``M-brane intersections from worldvolume superalgebras,''
  Phys.\ Lett.\  B {\bf 421}, 109 (1998)
  [arXiv:hep-th/9711043].
}

\lref\BaggerJR{
  J.~Bagger and N.~Lambert,
  ``Gauge Symmetry and Supersymmetry of Multiple M2-Branes,''
  arXiv:0711.0955 [hep-th].
}

\lref\GomisUV{
  J.~Gomis, G.~Milanesi and J.~G.~Russo,
  ``Bagger-Lambert Theory for General Lie Algebras,''
  arXiv:0805.1012 [hep-th].
}

\lref\BasuED{
  A.~Basu and J.~A.~Harvey,
  ``The M2-M5 brane system and a generalized Nahm's equation,''
  Nucl.\ Phys.\  B {\bf 713}, 136 (2005)
  [arXiv:hep-th/0412310].
}

\lref\BaggerSK{
  J.~Bagger and N.~Lambert,
  ``Modeling multiple M2's,''
  Phys.\ Rev.\  D {\bf 75}, 045020 (2007)
  [arXiv:hep-th/0611108].
  }
\lref\BaggerVI{
  J.~Bagger and N.~Lambert,
  ``Comments On Multiple M2-branes,''
  JHEP {\bf 0802}, 105 (2008)
  [arXiv:0712.3738 [hep-th]].
}
\lref\GustavssonVU{
  A.~Gustavsson,
  arXiv:0709.1260 [hep-th].
}

\lref\SchwarzYJ{
  J.~H.~Schwarz,
  ``Superconformal Chern-Simons theories,''
  JHEP {\bf 0411}, 078 (2004)
  [arXiv:hep-th/0411077].
}
\lref\ConstableAC{
  N.~R.~Constable, R.~C.~Myers and O.~Tafjord,
  ``The noncommutative bion core,''
  Phys.\ Rev.\  D {\bf 61}, 106009 (2000)
  [arXiv:hep-th/9911136].
}
\lref\PapadopoulosUQ{
  G.~Papadopoulos and P.~K.~Townsend,
  ``Intersecting M-branes,''
  Phys.\ Lett.\  B {\bf 380}, 273 (1996)
  [arXiv:hep-th/9603087].
}
\lref\TownsendAF{
  P.~K.~Townsend,
  ``D-branes from M-branes,''
  Phys.\ Lett.\  B {\bf 373}, 68 (1996)
  [arXiv:hep-th/9512062].
}

\lref\TseytlinBH{
  A.~A.~Tseytlin,
  ``Harmonic superpositions of M-branes,''
  Nucl.\ Phys.\  B {\bf 475}, 149 (1996)
  [arXiv:hep-th/9604035].
}
\lref\StromingerAC{
  A.~Strominger,
  ``Open p-branes,''
  Phys.\ Lett.\  B {\bf 383}, 44 (1996)
  [arXiv:hep-th/9512059].
}
\lref\GauntlettSS{
  J.~P.~Gauntlett, J.~Gomis and P.~K.~Townsend,
  ``BPS bounds for worldvolume branes,''
  JHEP {\bf 9801}, 003 (1998)
  [arXiv:hep-th/9711205].
}
\lref\MyersPS{
  R.~C.~Myers,
  ``Dielectric-branes,''
  JHEP {\bf 9912}, 022 (1999)
  [arXiv:hep-th/9910053].
}

\lref\WittenMH{
  E.~Witten and D.~I.~Olive,
  ``Supersymmetry Algebras That Include Topological Charges,''
  Phys.\ Lett.\  B {\bf 78}, 97 (1978).
}
\lref\AlvarezGaumeMV{
  L.~Alvarez-Gaume and S.~F.~Hassan,
  ``Introduction to S-duality in N = 2 supersymmetric gauge theories: A
  pedagogical review of the work of Seiberg and Witten,''
  Fortsch.\ Phys.\  {\bf 45}, 159 (1997)
  [arXiv:hep-th/9701069].
}
\lref\GukovJK{
  S.~Gukov and E.~Witten,
  ``Gauge theory, ramification, and the geometric langlands program,''
  arXiv:hep-th/0612073.
}
\lref\GomisFI{
  J.~Gomis and S.~Matsuura,
  ``Bubbling surface operators and S-duality,''
  JHEP {\bf 0706}, 025 (2007)
  [arXiv:0704.1657 [hep-th]].
}

\lref\GustavssonBF{
  A.~Gustavsson,
  ``One-loop corrections to Bagger-Lambert theory,''
  arXiv:0805.4443 [hep-th].
}
\lref\FigueroaOFarrillZM{
  J.~Figueroa-O'Farrill, P.~de Medeiros and E.~Mendez-Escobar,
  ``Lorentzian Lie 3-algebras and their Bagger-Lambert moduli space,''
  arXiv:0805.4363 [hep-th].
}
\lref\LinQP{
  H.~Lin,
  ``Kac-Moody Extensions of 3-Algebras and M2-branes,''
  arXiv:0805.4003 [hep-th].
}
\lref\BanerjeePD{
  S.~Banerjee and A.~Sen,
  ``Interpreting the M2-brane Action,''
  arXiv:0805.3930 [hep-th].
}
\lref\HosomichiJD{
  K.~Hosomichi, K.~M.~Lee, S.~Lee, S.~Lee and J.~Park,
  ``N=4 Superconformal Chern-Simons Theories with Hyper and Twisted Hyper
  Multiplets,''
  arXiv:0805.3662 [hep-th].
}
\lref\LiEZ{
  M.~Li and T.~Wang,
  ``M2-branes Coupled to Antisymmetric Fluxes,''
  arXiv:0805.3427 [hep-th].
}
\lref\JeonBX{
  I.~Jeon, J.~Kim, N.~Kim, S.~W.~Kim and J.~H.~Park,
  ``Classification of the BPS states in Bagger-Lambert Theory,''
  arXiv:0805.3236 [hep-th].
}
\lref\SongBI{
  Y.~Song,
  ``Mass Deformation of the Multiple M2 Branes Theory,''
  arXiv:0805.3193 [hep-th].
}
\lref\KrishnanZM{
  C.~Krishnan and C.~Maccaferri,
  ``Membranes on Calibrations,''
  arXiv:0805.3125 [hep-th].
}
\lref\HoVE{
  P.~M.~Ho, Y.~Imamura, Y.~Matsuo and S.~Shiba,
  ``M5-brane in three-form flux and multiple M2-branes,''
  arXiv:0805.2898 [hep-th].
}
\lref\FujiYJ{
  H.~Fuji, S.~Terashima and M.~Yamazaki,
  ``A New N=4 Membrane Action via Orbifold,''
  arXiv:0805.1997 [hep-th].
}
\lref\HonmaUN{
  Y.~Honma, S.~Iso, Y.~Sumitomo and S.~Zhang,
  ``Janus field theories from multiple M2 branes,''
  arXiv:0805.1895 [hep-th].
}
\lref\MorozovRC{
  A.~Morozov,
  ``From Simplified BLG Action to the First-Quantized M-Theory,''
  arXiv:0805.1703 [hep-th].
}
\lref\HoEI{
  P.~M.~Ho, Y.~Imamura and Y.~Matsuo,
  ``M2 to D2 revisited,''
  arXiv:0805.1202 [hep-th].
}
\lref\BenvenutiBT{
  S.~Benvenuti, D.~Rodriguez-Gomez, E.~Tonni and H.~Verlinde,
  ``N=8 superconformal gauge theories and M2 branes,''
  arXiv:0805.1087 [hep-th].
}
\lref\GomisUV{
  J.~Gomis, G.~Milanesi and J.~G.~Russo,
  ``Bagger-Lambert Theory for General Lie Algebras,''
  arXiv:0805.1012 [hep-th].
}
\lref\HoNN{
  P.~M.~Ho and Y.~Matsuo,
  ``M5 from M2,''
  arXiv:0804.3629 [hep-th].
}
\lref\PapadopoulosGH{
  G.~Papadopoulos,
  ``On the structure of k-Lie algebras,''
  arXiv:0804.3567 [hep-th].
}
\lref\GauntlettUF{
  J.~P.~Gauntlett and J.~B.~Gutowski,
  ``Constraining Maximally Supersymmetric Membrane Actions,''
  arXiv:0804.3078 [hep-th].
}

\lref\PapadopoulosSK{
  G.~Papadopoulos,
  ``M2-branes, 3-Lie Algebras and Plucker relations,''
  JHEP {\bf 0805}, 054 (2008)
  [arXiv:0804.2662 [hep-th]].
}
\lref\HosomichiQK{
  K.~Hosomichi, K.~M.~Lee and S.~Lee,
  ``Mass-Deformed Bagger-Lambert Theory and its BPS Objects,''
  arXiv:0804.2519 [hep-th].
}
\lref\BergshoeffCZ{
  E.~A.~Bergshoeff, M.~de Roo and O.~Hohm,
  ``Multiple M2-branes and the Embedding Tensor,''
  arXiv:0804.2201 [hep-th].
}
\lref\GomisCV{
  J.~Gomis, A.~J.~Salim and F.~Passerini,
  ``Matrix Theory of Type IIB Plane Wave from Membranes,''
  arXiv:0804.2186 [hep-th].
}
\lref\HoBN{
  P.~M.~Ho, R.~C.~Hou and Y.~Matsuo,
  ``Lie 3-Algebra and Multiple M2-branes,''
  arXiv:0804.2110 [hep-th].
}
\lref\GranVI{
  U.~Gran, B.~E.~W.~Nilsson and C.~Petersson,
  ``On relating multiple M2 and D2-branes,''
  arXiv:0804.1784 [hep-th].
}
\lref\DistlerMK{
  J.~Distler, S.~Mukhi, C.~Papageorgakis and M.~Van Raamsdonk,
  ``M2-branes on M-folds,''
  JHEP {\bf 0805}, 038 (2008)
  [arXiv:0804.1256 [hep-th]].
}
\lref\LambertET{
  N.~Lambert and D.~Tong,
  ``Membranes on an Orbifold,''
  arXiv:0804.1114 [hep-th].
}
\lref\MorozovCB{
  A.~Morozov,
  ``On the Problem of Multiple M2 Branes,''
  JHEP {\bf 0805}, 076 (2008)
  [arXiv:0804.0913 [hep-th]].
}
\lref\VanRaamsdonkFT{
  M.~Van Raamsdonk,
  ``Comments on the Bagger-Lambert theory and multiple M2-branes,''
  arXiv:0803.3803 [hep-th].
}
\lref\BermanBE{
  D.~S.~Berman, L.~C.~Tadrowski and D.~C.~Thompson,
  ``Aspects of Multiple Membranes,''
  arXiv:0803.3611 [hep-th].
}
\lref\BandresVF{
  M.~A.~Bandres, A.~E.~Lipstein and J.~H.~Schwarz,
  ``N = 8 Superconformal Chern--Simons Theories,''
  JHEP {\bf 0805}, 025 (2008)
  [arXiv:0803.3242 [hep-th]].
}
\lref\MukhiUX{
  S.~Mukhi and C.~Papageorgakis,
  ``M2 to D2,''
  JHEP {\bf 0805}, 085 (2008)
  [arXiv:0803.3218 [hep-th]].
}
\lref\GustavssonDY{
  A.~Gustavsson,
  ``Selfdual strings and loop space Nahm equations,''
  JHEP {\bf 0804}, 083 (2008)
  [arXiv:0802.3456 [hep-th]].
}



\Title{
\vbox{\baselineskip12pt
\hbox{}}}
{\vbox{\centerline{M2-Brane  Superalgebra from Bagger-Lambert Theory}}}

\smallskip
\smallskip

\vskip-5pt
\centerline{Filippo Passerini}
\smallskip

\smallskip
\smallskip
\bigskip
\medskip
\centerline{\it Perimeter Institute for Theoretical Physics}
\centerline{\it Waterloo, Ontario N2L 2Y5, Canada}
\bigskip
\centerline{\it and}
\bigskip
\centerline{\it Department of Physics and Astronomy}
\centerline{\it University of Waterloo,  Ontario N2L 3G1, Canada}
\vskip .1in

\footnote{${}^{}$}{\tt fpasserini@perimeterinstitute.ca}

\vskip .2in
\centerline{\bf{Abstract}}
It is known that the M2-brane worldvolume superalgebra includes  two p-form central charges  that encode  the M-theory intersections  involving M2-branes. In this paper we show by explicit computation that the Bagger-Lambert Lagrangian realizes the M2-brane superalgebra,  including also the central extensions. Solitons of the Bagger-Lambert theory, that are interpreted as worldvolume realizations of intersecting branes, are  shown to  saturate a BPS-bound  given in terms of the corresponding central charge.      
\medskip
\medskip

\Date{06/2008}


\newsec{Introduction and Discussion}

Brane   intersections can be described as  solitons of the  worldvolume theory of one of the constituents of the intersecting system \CallanKZ \GibbonsXZ . In particular, quarter-BPS intersections appear on the worldvolume as half-BPS solitons and the spacetime interpretation relies on the fact that the worldvolume scalars  encode the brane embedding.  

Many M-branes systems in M-theory have been studied  using this approach.  For instance, a stack  of M2-branes ending on an M5-brane   is associated to a self-dual string soliton on the  M5-brane worldvolume  \HoweUE\    and the M5-M5 intersection can be described as a  3-brane vortex  on the worldvolume of  one of the M5-branes   \HoweET.  In a similar way, the   M2-M2 intersection can be described as a 0-brane vortex on the worldvolume of one of the M2-branes \CallanKZ\GauntlettSS.  All these examples mentioned are the worldvolume realization of previously studied quarter-BPS intersecting systems \PapadopoulosUQ\TownsendAF\TseytlinBH\StromingerAC.

In a recent paper \BaggerJR, Bagger and Lambert have  proposed a Lagrangian to describe the low energy dynamics of a stack of coincident M2-branes (see also the work  by Gustavsson \GustavssonVU).  Their model, that incorporates insights from previous papers \BaggerSK\SchwarzYJ, includes half-BPS fuzzy 3-sphere solitons. This solutions were argued by Basu and Harvey \BasuED\   to provide the M2-branes worldvolume description of the multiple M2-branes ending on an M5-brane,  generalizing a similar mechanism studied for the D1-D3 system \ConstableAC.  The  Bagger-Lambert theory is a 3-dimensional ${\cal N}=8$ supersymmetric field theory, based on a novel algebraic structure, dubbed 3-algebra.  Explicit examples of 3-algebras has been recently constructed  in \GomisUV\BenvenutiBT\HoEI\ starting  from ordinary Lie algebras and  considering a Lorentzian scalar product (see also \LinQP\FigueroaOFarrillZM). The  fact that the scalar  product is not positive-definite permit to avoid a no-go theorem discussed in \PapadopoulosSK \GauntlettUF.  Other algebraic structures have been considered in \GustavssonDY\MorozovCB\HoBN\PapadopoulosGH\HoNN. The Bagger-Lambert     theory was shown to be conformal invariant in \BandresVF\  and the moduli space was discussed in \BaggerVI\VanRaamsdonkFT\LambertET\DistlerMK\JeonBX. The maximally supersymmetric deformation of the theory was constructed in \GomisCV\HosomichiQK (see  also \SongBI), for other deformations of the theory see \BermanBE\HonmaUN\FujiYJ\LiEZ. In \BergshoeffCZ\ the Bagger-Lambert theory is derived applying the embedding tensor methods and in \MukhiUX\GranVI\  the reduction to the theory of multiple D2-branes is discussed.  Other recent developments are in \HoVE\KrishnanZM\HosomichiJD\GustavssonBF\MorozovRC.     

It was shown in \BergshoeffBH\  that the  spacetime interpretation of the worldvolume solitons can be deduced also from the  worldvolume supersymmetry algebra. For the case of the M2-brane the worldvolume supersymmetry algebra is given by the maximal central extension of the 3-dimensional ${\cal N}=8$ super-Poincare algebra  \BergshoeffBH. The   anticommutator is given by     
\eqn\introqq{\eqalign{\{Q_{\hat{\alpha}}^p, Q_{\hat{\beta}}^q\}=-2P_{\mu}(\hat{\gamma}^{\mu}\hat{\gamma}^0)_{\hat{\alpha}\hat{\beta}}\delta^{pq} +Z^{[pq]}\varepsilon_{\hat{\alpha}\hat{\beta}}+Z^{(pq)}_\mu(\hat{\gamma}^{\mu}\hat{\gamma}^0)_{\hat{\alpha}\hat{\beta}}}}
where $Q_{\hat{\alpha}}^p$  are the eight 3-dimensional Majorana spinor supercharges and $Z^{[pq]}$,$Z^{(pq)}_\mu$ are the 0-form and the 1-form worldvolume central charges. $p,q=1,\ldots 8$ are the indices of the SO(8) automorphism group    and the supercharges transform  as chiral spinors of SO(8).  Due to the triality relation of SO(8), we can  consider the supercharges  to transform in the vector representation of SO(8) and thus we can interpret the automorphism group SO(8) as the rotation group in the eight directions transverse to the M2-branes.    The 0-form  $Z^{[pq]}$ is in the {\bf 28} representation of  SO(8) and it can be thought as   a 2-form in the transverse space.  This central charge is  associated  with  M2-branes that are intersecting the original M2-branes along the time direction,  a quarter-BPS  system studied in \PapadopoulosUQ.  The 1-form  $Z_\mu^{(pq)}$ is in the ${\bf 35}^+$ of SO(8) and it is a self-dual 4-form in the transverse space. This implies that the  1-form charge is associated to the quarter-BPS M2-M5  system \StromingerAC\TownsendAF.  

We have seen that the M2-brane superalgebra, correctely incorporates all the possible  quarter-BPS intersections  between the M2-branes and the other M-branes of M-theory.\foot{In the worldvolume description, these intersections are half-BPS solitons.}  This implies that a  complete M2-branes worldvolume theory should realize the M2-brane superalgebra \introqq, including also the central charges.  
      
In this paper, we verify by explicit computation that the Bagger-Lambert theory does realize the M2-brane superalgebra \introqq. The central charges that we obtain  are  given by

\eqn\centralintro{\eqalign {&Z^{[pq]}= -\int d^2\sigma\partial_i\hbox{Tr}( X^I,D_j X^{J})\varepsilon^{ij}(\gamma^{IJ})^{pq}\cr
&Z^{(pq)}_\mu=-{1\over 12} \int d^2\sigma\partial_i\hbox{Tr}( X^I,[ X^J,X^K,X^M]) \varepsilon^{0i}{}_\mu(\gamma^{IJKM})^{pq} .}}  
We note that  the 0-form charge  $Z^{[pq]}$ is  the natural generalization of the  charge computed in \GauntlettSS\ using the BPS-bound for the vortex  solution in the single M2-brane theory.  The 1-form instead relies on the non-abelian nature of  the scalar fields in the Bagger-Lambert theory  and it vanishes in the limit  where the stack of multiple  M2-branes  reduces to a single M2-brane.  This is consistent with the fact that the M2-M5 intersection cannot be seen on the worldvolume of a single M2-brane. Indeed, given   an intersection between branes with different dimensions, the worldvolume  description of the system using the worldvolume of the lower dimensional brane is usually based on non-abelianity  \ConstableAC\MyersPS.      

We show  that a vortex solution excites the 0-form central charge and the Basu-Harvey solution excites the  1-form central charge, in agreement with the interpretation of this solitons  as the quarter-BPS M2-M2 intersection and the quarter-BPS M2-M5 intersection.  The energy of this configurations is bounded below by the value of the corresponding central charge and the bound is saturated when the solitons  are half-BPS. This is in  agreement with the structure of the M2-brane superalgebra \centralintro. 

The rest of the paper is organized as follows. In section 2 we briefly review the Bagger-Lambert theory and we write down  the supercurrent associated to the supersymmetry  of the Lagrangian.  This enables us to  express the supercharges in terms of the fields of the  theory.  In section 3 we use the field theory realization of the supercharges to compute the  central charges. In section 4 we analyze the vortex and the Basu-Harvey solitons and show that they are associated to  central charges, in agreement with  the interpretation of this solutions as the worldvolume realization of intersecting systems.  Appendix A  summarizes our notation and appendix B includes the proof of the conservation of the supercurrent.  Appendix C contains technical details of the computation for the central charges.        

\newsec{The Bagger-Lambert  Theory}
\subsec{The Lagrangian}
We start reviewing the Lagrangian  proposed  by Bagger and Lambert \BaggerJR\  as the low energy  effective theory for multiple coincident M2-branes.  In this model,  the transverse fluctuations of the  membranes are described by eight scalar fields $X^I$, where $I=3,\ldots 11$  and the  eight $Spin(1,2)$ worldvolume  fermions are collected together in the  spinor field  $\Psi$. The $\Psi$ is an  11-dimensional Majorana spinor satisfying the condition $\Gamma_{012}\Psi=-\Psi$ and thus it has sixteen independent real components\foot{We summarize  our conventions in appendix $A$. }. 

These fields are valued in a 3-algebra ${\cal A}$ \BaggerJR (see also \GustavssonVU),  i.e. $X^I=X^I_aT^a$ and   $\Psi=\Psi_aT^a$ where  $T^a$, $a=1,\ldots, \hbox{dim}{\cal A}$ are the generators of ${\cal A}$.  The 3-algebra is endowed with a 3-product 
\eqn\prod{
[T^a,T^b,T^c]=f^{abc}_{~~~~~d}T^d} 
where the structure constants satisfy the fundamental identity
\eqn\fundid{f^{efg}{}_{d}f^{abc}{}_{g}=f^{efa}{}_{g}f^{bcg}{}_{d}
+f^{efb}{}_{g}f^{cag}{}_{d}+f^{efc}{}_{g}f^{abg}{}_{d}.}
The  3-algebra construction includes also a bilinear and non-degenerate scalar product Tr$(\cdot, \cdot)$ that defines a non-degenarate metric $h^{ab}$ 
\eqn\metr{h^{ab}\equiv\hbox{Tr}(T^a,T^b)}
used to manipulate the algebra indices.  The structure constants $f^{abcd}$ are assumed to be totally antisymmetric in the indices. 
   
The   Bagger-Lambert  theory includes also a non-propagating gauge vector field $A_{\mu ab}$ where $\mu=0,1,2$ denotes the worldvolume coordinates. The dynamics is controlled by the   Lagrangian  
\eqn\lag{\eqalign{
{\cal L} =& -{1\over2}(D_\mu X^{aI})(D^\mu X^{I}_{a})
+{i\over 2}\bar\Psi^a\Gamma^\mu D_\mu \Psi_a
+{i\over4}\bar\Psi_b\Gamma_{IJ}X^I_cX^J_d\Psi_a f^{abcd}\cr
& - V+{1\over2}\varepsilon^{\mu\nu\lambda}(f^{abcd}A_{\mu
ab}\partial_\nu A_{\lambda cd} +{2\over 3}f^{cda}{}_gf^{efgb}
A_{\mu ab}A_{\nu cd}A_{\lambda ef})}}
where $V$ is the potential
\eqn\potential{\eqalign{V  = {1\over 12}f^{abcd}f^{efg}{}_d
X^I_aX^J_bX^K_cX^I_eX^J_fX^K_g 
 = {1\over2\cdot 3!}\hbox{Tr}([X^I,X^J,X^K],[X^I,X^J,X^K])}}
 and  the covariant derivative of  a  field $\Phi$ is defined by 
 \eqn\covari{(D_{\mu} \Phi)_a = \partial_{\mu} \Phi_a - \tilde {A}_{\mu}{}^b{}_a \Phi_b}
 where $\tilde {A}_{\mu}{}^b{}_a\equiv f^{cdb}{}_a A_{\mu cd}$.  The \lag\ is  invariant under the gauge transformations
 \eqn\gauge{\eqalign{\delta X^I_a =&\tilde{\Lambda}^b{}_a X^I_b \cr
\delta \Psi_a =& \tilde{\Lambda}^b{}_a\Psi_b\cr
 \delta\tilde{A}_{\mu}{}^b{}_a =& \partial_\mu\tilde{\Lambda}^b{}_a-\tilde{\Lambda}^b{}_c\tilde{A}_{\mu}{}^c{}_a+\tilde{A}_{\mu}{}^b{}_c\tilde{\Lambda}^c{}_a }}
 where $\tilde{\Lambda}{}^b{}_a\equiv f^{cdb}{}_a \Lambda_{cd}$ and  $\Lambda_{cd}$ is the gauge parameter.  The Lagrangian \lag\ is also invariant under the   following supersymmetry variations
\eqn\susyb{\eqalign{\delta_\epsilon X^I_a =& i\bar\epsilon\Gamma^I\Psi_a\cr
\delta_\epsilon \Psi_a =& D_\mu X^I_a\Gamma^\mu \Gamma^I\epsilon -{1\over 6}
X^I_bX^J_cX^K_d f^{bcd}{}_{a}\Gamma^{IJK}\epsilon \cr
 \delta_\epsilon\tilde{A}_{\mu}{}^b{}_a =& i\bar\epsilon
\Gamma_\mu\Gamma_IX^I_c\Psi_d f^{cdb}{}_{a}}}
where the supersymmetry parameter $\epsilon$  satisfies  $\Gamma_{012}\epsilon=\epsilon$. The equations of motion are
\eqn\eomf{\eqalign{\Gamma^\mu D_\mu\Psi_a
+{1\over 2}\Gamma_{IJ}X^I_cX^J_d\Psi_bf^{cdb}{}_{a}=&0\cr
 D^2X^I_a-{i\over2}\bar\Psi_c\Gamma^I_{ J}X^J_d\Psi_b f^{cdb}{}_a
   -{\partial V\over \partial X^{Ia}}    =& 0 \cr
\tilde F_{\mu\nu}{}^b{}_a
  +\varepsilon_{\mu\nu\lambda}(X^J_cD^\lambda X^J_d
+{i\over2}\bar\Psi_c\Gamma^\lambda\Psi_d )f^{cdb}{}_{a}  =& 0}}
where 
\eqn\ftmunu{\tilde{F}_{\mu\nu}{}^b{}_a  =\partial_\nu \tilde{A}_{\mu}{}^b{}_a  -
\partial_\mu \tilde{A}_{\nu}{}^b{}_a-\tilde{A}_{\mu}{}^b{}_c\tilde{A}_{\nu}{}^c{}_a
+ \tilde{A}_{\nu}{}^b{}_c \tilde{A}_{\mu}{}^c{}_a.}

The  stress-energy tensor $T_{\mu\nu}$ can be computed in the usual way coupling  the  Bagger-Lambert theory to an external worldvolume  metric  and  looking at the variation of the action for an infinitesimal change of the metric. 
In the case where the fermions are set to zero, it results 
\eqn\tmunu{\eqalign{T_{\mu\nu}=D_\mu X_a^{I}D_\nu X^{a I}-\eta_{\mu\nu}\left({1\over2}D_\rho X^{aI}D^\rho X^{I}_{a}+V\right).}} 
We note that the Chern-Simons like term in \lag\ does not contribute to the stress-energy tensor.  This is because this term is topological and does not depend on the worldvolume metric.
\subsec{Supercharges}
Given the invariance of the Lagrangian \lag\  under the supersymmetry variations  \susyb,  the  Noether theorem implies the existence of a conserved supercurrent  $J^\mu$ given by 
\eqn\jz{J^\mu=-D_\nu X_a^I\Gamma^\nu\Gamma^I\Gamma^\mu \Psi^a-{1\over 6} X_a^IX_b^JX_c^K f^{abcd}\Gamma^{IJK}\Gamma^\mu\Psi_d.}
In Appendix $B$ we show that $\partial_\mu J^\mu=0$.  The supercharge is thus  the integral over the spatial worldvolume coordinates of the timelike component of the supercurrent, i.e.
\eqn\scharge{\eqalign{Q=&\int d^2 \sigma J^{0}\cr
=&-\int d^2 \sigma(D_\nu X_a^I\Gamma^\nu\Gamma^I\Gamma^0 \Psi^a+{1\over 6} X_a^IX_b^JX_c^K f^{abcd}\Gamma^{IJK}\Gamma^0\Psi_d).}}

Given that the mass dimensions of the fields in the Bagger-Lambert theory are $[X]={1\over 2}$ and  $[A]=[\Psi]=1$,   
it follows that $J^0$ has  mass dimension $[J^0]={5\over 2}$.  This  gives $[Q]={1\over 2}$,  that is the right mass dimension for the supercharge. It is easy to check that the two terms on the right hand side of \jz\ are the only gauge invariant combinations of fields with the right mass dimension and  with an uncontracted spinorial index.

The supercharge $Q$ is the generator of the supersymmetry transformation, that means that  the supersymmetry variation of a field $\Phi$ is given by $\delta_{\epsilon}\Phi=[\bar{\epsilon}Q,\Phi]$.  More in details, for Grassman-even and Grassman-odd fields $\Phi_E$ and  $\Phi_O$   we have  
\eqn\var{\delta_{\epsilon}\Phi_E=\bar{\epsilon}_{\alpha}[Q^{\alpha},\Phi_E]\qquad \delta_{\epsilon}\Phi_O^\beta=\bar{\epsilon}_{\alpha}\{Q^{\alpha},\Phi_O^\beta\}} 
where we have explicitly shown the 11-dimension spinorial indices $\alpha$ and $\beta$.  Using the canonical commutation relations, one can show that the \var\  reproduce the supersymmetry variations of the Bagger-Lambert theory   \susyb.

\newsec{Central Charges}
In this section, we show that the supersymmetry algebra of the Bagger-Lambert theory  includes  two  central charge forms, as  expected for a theory describing M2-branes.   These central extensions  are computed here explicitly using the  field  realization of the supercharge $Q$  given in \scharge \WittenMH.\foot{For a review, see for instance \AlvarezGaumeMV.} In details,  we consider the relation
\eqn\rel{\bar{\epsilon}_{\alpha}\{Q^{\alpha},Q^{\beta}\}=\int d^2 \sigma\bar{\epsilon}_{\alpha}\{Q^{\alpha},J^{0\beta}(\sigma)\}=\int d^2 \sigma\delta_{\epsilon}J^{0\beta}(\sigma)}
where in the last step we used the second of the equations  \var.  The  supersymmetry variation of the zeroth component of the supercurrent  $\delta_{\epsilon}J^{0}$ is computed in the Appendix $C$.  For the case where the spinors $\Psi$ are set to zero, it is given by   
\eqn\cvardue{\eqalign{\delta_{\epsilon}J^{0}=-2T^{0}{}_{\mu}\Gamma^{\mu}\epsilon  -\partial_i( X_a^I D_j X^{aJ}\varepsilon^{ij}\Gamma^{IJ}\epsilon)-{1\over 12} \partial_i( X_a^I X_b^JX_c^KX_d^M f^{bcda}\varepsilon^{0i}{}_\mu\Gamma^{IJKM}\Gamma^{\mu}\epsilon)}}
where $i=1,2$ labels the spatial worldvolume coordinates. From the expression \cvardue\ and the relation  \rel\ we get 
\eqn\aceleven{\eqalign{\{Q^\alpha, Q^\beta\}=&-2P_{\mu}(\Gamma^{\mu}\Gamma^0)^{\alpha\beta}  -\int d^2\sigma\partial_i( X_a^I D_j X^{aJ}\varepsilon^{ij})(\Gamma^{IJ}\Gamma^0)^{\alpha\beta}\cr &-{1\over 12} \int d^2\sigma\partial_i( X_a^I X_b^JX_c^KX_d^M f^{bcda}\varepsilon^{0i}{}_\mu)(\Gamma^{IJKM}\Gamma^{\mu}\Gamma^0)^{\alpha\beta}}}
where $P^\mu$ is the energy momentum vector defined as $P^\mu=\int d^2 \sigma T^{0\mu}$.
\subsec{Spinors Decomposition}
In order to better analyze the structure of the ${\cal N}=8$  superalgebra, we need to write  the anticommutator \aceleven\  in terms of  3-dimensional spinors.  To this end, we decompose  the $Spin(1,10)$ Dirac matrices in  terms of $Spin(1,2)\otimes Spin(8)$ Dirac matrices.  In details, we take
\eqn\gammadec{\Gamma^\mu=\hat{\gamma}^\mu\otimes\bar{\gamma}_9\qquad\hbox{and}\qquad\Gamma^I=1\otimes\bar{\gamma}^I}
where 
\eqn\algebra{\{\hat{\gamma}^\mu,\hat{\gamma}^\nu\}=2\eta^{\mu\nu},\qquad \{\bar{\gamma}^I,\bar{\gamma}^J\}=2\delta^{IJ},\qquad\qquad\bar{\gamma}_9=\bar{\gamma}^3\ldots\bar{\gamma}^{10}}
and it is easy to check that the matrices \gammadec\  satisfies the 11-dimensinal Clifford algebra. The  $\hat{\gamma}^\mu$ are $2\times 2$  real matrices. Explicitly
\eqn\gexp{\hat{\gamma}^0=i\sigma^2_{\hat{\alpha}\hat{\beta}}=\varepsilon_{\hat{\alpha}\hat{\beta}}\qquad\hat{\gamma}^1=\sigma^1_{\hat{\alpha}\hat{\beta}}\qquad\hat{\gamma}^2=\sigma^3_{\hat{\alpha}\hat{\beta}}}
where the $\sigma$'s are Pauli matrices and $\hat{\alpha},\hat{\beta}=1,2$ are 3-dimensional spinorial indices.  The $\bar{\gamma}^I$ are $16\times16$ real matrices given by 
\eqn\gammabar{\bar{\gamma}^I=\left(\eqalign{0\qquad \gamma^I_{\dot{p}p}\cr \gamma^I_{q\dot{q}}\qquad 0}\right)}
where $(\gamma^{I}_{p\dot{p}})^T=\gamma^{I}_{\dot{p}p}$ are $8\times 8$ real gamma matrices satisfying 
\eqn\clifdot{\eqalign{\gamma^{I}_{p\dot{p}}\gamma^{J}_{\dot{p}q}+\gamma^{J}_{p\dot{p}}\gamma^{I}_{\dot{p}q}=2\delta^{IJ}\delta_{pq}\qquad \gamma^{I}_{\dot{p}p}\gamma^{J}_{p\dot{q}}+\gamma^{J}_{\dot{p}p}\gamma^{I}_{p\dot{q}}=2\delta^{IJ}\delta_{\dot{p}\dot{q}}.}} 
 Given that $\Gamma_{012}=-\hat{\gamma}_{012}\otimes\bar{\gamma}_9=-1\otimes\bar{\gamma}_9$,    spinors with definite $\Gamma_{012}$ chirality, have a definite $\bar{\gamma}_9$ chirality.\foot{In this representation  $\bar{\gamma}_9=\left(\eqalign{1&\qquad 0\cr 0& \quad -1}\right)$.}  

Using the matrices decomposition just described and the fact that $\Gamma_{012}Q=Q$, the equation \aceleven\ can be written as  
\eqn\avthree{\eqalign{\{Q_{\hat{\alpha}}^p, Q_{\hat{\beta}}^q\}=&-2P_{\mu}(\hat{\gamma}^{\mu}\hat{\gamma}^0)_{\hat{\alpha}\hat{\beta}}\delta^{pq}  -\int d^2\sigma\partial_i( X_a^I D_j X^{aJ}\varepsilon^{ij})(\gamma^{IJ})^{pq}\varepsilon_{\hat{\alpha}\hat{\beta}}\cr &-{1\over 12} \int d^2\sigma\partial_i( X_a^I X_b^JX_c^KX_d^M f^{bcda}\varepsilon^{0i}{}_\mu)(\gamma^{IJKM})^{pq} (\hat{\gamma}^{\mu}\hat{\gamma}^0)_{\hat{\alpha}\hat{\beta}}}}
where $(\gamma^{IJ})_{pq}=\gamma^{[I}_{p\dot{r}}\gamma^{J]}_{\dot{r}q}$ and $(\gamma^{IJKM})_{pq}=\gamma^{[I}_{p\dot{r}}\gamma^{J}_{\dot{r}r}\gamma^{K}_{r\dot{t}}\gamma^{M]}_{\dot{t}q}$.

Thus, we conclude that the Bagger-Lambert Lagrangian realizes the centrally  extended 3-dimensional  ${\cal N}=8$ superalgebra 
\eqn\cvardech{\eqalign{\{Q_{\hat{\alpha}}^p, Q_{\hat{\beta}}^q\}=-2P_{\mu}(\hat{\gamma}^{\mu}\hat{\gamma}^0)_{\hat{\alpha}\hat{\beta}}\delta^{pq} +Z^{[pq]}\varepsilon_{\hat{\alpha}\hat{\beta}}+Z^{(pq)}_\mu(\hat{\gamma}^{\mu}\hat{\gamma}^0)_{\hat{\alpha}\hat{\beta}}}}
where the central charges are given by
\eqn\central{\eqalign {&Z^{[pq]}= -\int d^2\sigma\partial_i\hbox{Tr}( X^I,D_j X^{J})\varepsilon^{ij}(\gamma^{IJ})^{pq}\cr
&Z^{(pq)}_\mu=-{1\over 12} \int d^2\sigma\partial_i\hbox{Tr}( X^I,[ X^J,X^K,X^M]) \varepsilon^{0i}{}_\mu(\gamma^{IJKM})^{pq} .}}

Using the property $(\gamma^{IJ})^{qp}=-(\gamma^{IJ})^{pq}$,   $(\gamma^{IJKM})^{qp}= (\gamma^{IJKM})^{pq}$  it follows  that the 0-form central charge is antisymmetric in $p,q$ indices and the 1-form central charge is symmetric in  $p,q$.  Given that   $\varepsilon_{\hat{\alpha}\hat{\beta}}=-\varepsilon_{\hat{\beta}\hat{\alpha}}$ and  $(\hat{\gamma}^{\mu}\hat{\gamma}^0)_{\hat{\alpha}\hat{\beta}}=(\hat{\gamma}^{\mu}\hat{\gamma}^0)_{\hat{\beta}\hat{\alpha}}$  the right hand side of the \central\  is correctly  symmetric under  the exchange   $(p,\hat{\alpha})\leftrightarrow (q,\hat{\beta})$.

The equations \central\  give the field realization of the central charges of the extended 3-dimensional  ${\cal N}=8$ superalgebra. They are boundary terms and  they are equal to zero for  field configurations that are non-singular and topologically trivial.  In the next section we will discuss half-BPS configurations  that  excite the central charges of the Bagger-Lambert theory.     

\newsec{Solitons of the Bagger-Lambert Theory}
\subsec{Vortices}
We consider  vortex  configurations \CallanKZ\GauntlettSS\ where only the scalars $X^3$, $X^4$ and the gauge vector $\tilde{A}_{\nu}{}^b{}_a$ are excited.  Given the interpretation of the Bagger-Lambert theory as  a  theory of coincident M2-branes, these configurations  describe two stacks  of membranes intersecting along the  time direction\foot{This is the  analog of the vortex like solution for ${\cal N}=4$ SYM describing a surface operator interpreted as the intersection D3$\cap$D3$=R^2$ \GukovJK\GomisFI.} 
\smallskip
\eqn\mtwomtwo{\matrix{\ \ &0&1&2&3&4&5&6&7&8&9&10\cr
M2&\hbox{X}&\hbox{X}&\hbox{X}&&&&&&\cr
M2&\hbox{X}&&&\hbox{X}&\hbox{X}&&&&&}}
\medskip
\noindent
It is convenient to introduce the complex worldvolume  coordinates $z$ and $\bar{z}$
\eqn\comz{z=\sigma^1+i\sigma^2\qquad \bar{z}=\sigma^1-i\sigma^2}
and the complex scalars $\Phi$ and $\bar{\Phi}$  \eqn\comphi{\Phi={1\over 2}(X_3-iX_4)\qquad \bar{\Phi}={1\over 2}(X_3+iX_4).}
Thus, considering  a configuration where only  $\Phi$,  $\bar{\Phi}$ and $\tilde{A}_{\mu}{}^b{}_a $  are switched on, and such that $D_0 \Phi=D_0 \bar{\Phi}=0$,   the  BPS conditions that follow from the  supersymmetry variations \susyb\ reduce to 
\eqn\bpssol{\eqalign{ D_z \Phi\Gamma^z \Gamma^\Phi\epsilon + D_{\bar{z}}\Phi\Gamma^{\bar{z}} \Gamma^\Phi\epsilon 
+D_z \bar{\Phi}\Gamma^z \Gamma^{\bar{\Phi}}\epsilon + D_{\bar{z}}\bar{\Phi}\Gamma^{\bar{z}} \Gamma^{\bar{\Phi}}\epsilon =0}} 
where
\eqn\comgphi{\Gamma^\Phi=\Gamma^3+i\Gamma^4\qquad \Gamma^{\bar{\Phi}}=\Gamma^3-i\Gamma^4\qquad\Gamma^z=\Gamma^1+i\Gamma^2\qquad\Gamma^{\bar{z}}=\Gamma^1-i\Gamma^2.} 
For this configuration, the energy density is given by 
\eqn\hdens{{\cal H}=4\hbox{Tr}(D_z\Phi , D_{\bar{z}}\bar{\Phi})+4\hbox{Tr}(D_{\bar{z}}\Phi , D_z\bar{\Phi})={{\cal Z}^0\over 2}+8\hbox{Tr}(D_{\bar{z}}\Phi , D_z\bar{\Phi})}
where ${\cal Z}^0$ is the density of the 0-form central charge $Z^{[pq]}$ evaluated for this field configuration. Thus, considering a positive definite scalar product Tr$(\cdot, \cdot)$, it results ${\cal H}\ge{{\cal Z}^0\over 2}$ and the bound is saturated when 
\eqn\bbound{D_{\bar{z}}\Phi=D_{z}\bar{\Phi}=0.}  
When this last condition is satisfied,  it follows from the BPS equation \bpssol\ that the solution preserve the supersymmetries satisfying $\Gamma^{z}\Gamma^{\Phi}\epsilon=0$ or equivalently $\Gamma^{1234}\epsilon=\epsilon$.
Thus, for the case where the gauge field is equal to zero, i.e. $\tilde{A}_{\mu}{}^b{}_a =0$ the vortex configuration given by  \eqn\vartaz{\Phi={c_aT^a\over z}} 
where $c_a$ are arbitrary constants  is a half-BPS state.\foot{In the sense that it preserves half of the supersymmetries \susyb.} The singularity in $z=0$ excite  the 0-form central charge  $Z^{[pq]}$ \central, in agreement with the interpretation of the vortex solution as the brane intersection \mtwomtwo.

We now discuss  the case where also the gauge vector is excited and to analyze this configuration we use the 3-algebra constructed  in  \GomisUV. In this model,  the 3-algebra indices $a$ are split into $a=(0,\tilde{a},\varphi)$ and the structure constants are given by 
\eqn\sconst{f^{0 \tilde{a}\tilde{b}\tilde{c}}=f^{\varphi\tilde{a}\tilde{b}\tilde{c}}=C^{\tilde{a}\tilde{b}\tilde{c}}\ ,\qquad f^{0\varphi \tilde{a}\tilde{b}}=f^{\tilde{a}\tilde{b}\tilde{c}\tilde{d}}=0}
where $C^{\tilde{a}\tilde{b}\tilde{c}}$ are the structure constants of a compact semi-simple Lie algebra  satisfying  the usual Jacobi identity. 
The structure constants \sconst\ solve the fundamental identity \fundid\ and they are totally antisymmetric. Following   \GomisUV, we introduce null generators  on the 3-algebra
\eqn\nullt{T^{\pm}=\pm T^0+T^\phi}
and in this basis the structure constants become
\eqn\nullstruct{f^{+\tilde{a}\tilde{b}\tilde{c}}=2 C^{\tilde{a}\tilde{b}\tilde{c}}\ ,\qquad f_{-\tilde{a}\tilde{b}\tilde{c}}= C_{\tilde{a}\tilde{b}\tilde{c}} \ ,\qquad f^{-\tilde{a}\tilde{b}\tilde{c}}=f_{+\tilde{a}\tilde{b}\tilde{c}}=0.}
The  gauge vector $A_\mu^{ab}$  is decomposed as 
\eqn\vecdec{A^{\tilde{a}}_\mu \equiv A^{-\tilde{a}}_\mu\ ,\qquad  B^{\tilde{a}}_\mu\equiv{1\over 2} C^{\tilde{a}\tilde{b}\tilde{c}} A_{\mu \tilde{b}\tilde{c}}.}
We consider a  configuration where only the $\tilde{a}$ components of the scalar field  are excited, we call this field $\tilde{\Phi}$. Thus  
\eqn\scati{\tilde{\Phi}= {c_{\tilde{a}}T^{\tilde{a}}\over z}.}
Taking  $B_\mu=0$, the  equation \bpssol\ reduce   to

\eqn\bpssolti{\eqalign{ \tilde{D}_z \tilde{\Phi}\Gamma^z \Gamma^\Phi\epsilon +  \tilde{D}_{\bar{z}}\tilde{\Phi}\Gamma^{\bar{z}} \Gamma^\Phi\epsilon + \tilde{D}_z \bar{\tilde{\Phi}}\Gamma^z \Gamma^{\bar{\Phi}}\epsilon +  \tilde{D}_{\bar{z}}\bar{\tilde{\Phi}}\Gamma^{\bar{z}} \Gamma^{\bar{\Phi}}\epsilon =0}} 
 where 
\eqn\cdadj{\tilde{D}_\mu \tilde{\Phi}^{\tilde{a}} \equiv \partial_\mu \tilde{\Phi}^{\tilde{a}} 
+2 C^{\tilde{a}}_{~~\tilde{b}\tilde{c}}A_\mu^{\tilde{c}}\tilde{\Phi}^{\tilde{b}}}
is the covariant derivative for a field in the adjoint representation  of the Lie algebra  with structure constants $C^{\tilde{a}\tilde{b}\tilde{c}}$. The energy density now is  
\eqn\hdenst{{\cal H}=4\hbox{Tr}(\tilde{D}_z\tilde{\Phi} , \tilde{D}_{\bar{z}}\bar{\tilde{\Phi}})+4\hbox{Tr}(\tilde{D}_{\bar{z}}\tilde{\Phi} , \tilde{D}_z\bar{\tilde{\Phi}})={{\cal Z}^0\over 2}+8\hbox{Tr}(\tilde{D}_{\bar{z}}\tilde{\Phi} , \tilde{D}_z\bar{\tilde{\Phi}})} and given that \GomisUV\ $\hbox{Tr}(T^{\tilde{a}},T^{\tilde{b}})=\delta^{\tilde{a}\tilde{b}}$,  it results   ${\cal H}\ge{{\cal Z}^0\over 2}$. The ${\cal Z}^0$ is the 0-form central charge evaluated for this solution and the  BPS-bound is saturated when $\tilde{D}_{\bar{z}}\tilde{\Phi}=\tilde{D}_{z}\bar{\tilde{\Phi}}=0$. Thus, it follows that the  configuration where only $\tilde{\Phi}$ and ${\cal A}_\mu=A_{\tilde{a}\mu}T^{\tilde{a}}$ are excited, is half-BPS if
\eqn\commu{[\tilde{\Phi},{\cal A}_{\bar{z}}]=[\bar{\tilde{\Phi}},{\cal A}_{z}]=0}
where $[\cdot,\cdot]$ is the usual Lie commutator. Also in this case, the preserved supersymmetries satisfy $\Gamma^{z}\Gamma^{\Phi}\epsilon=0$ and this configuration excites the  the 0-form central charge.
This implies that with respect to the single M2-brane theory, the vortex solutions of the Bagger-Lambert theory includes  extra degrees of freedom,  given by the the  components of the  gauge vector that commute with the scalar fields.

\subsec{Basu-Harvey  Solitons}
To describe a stack of M2-branes ending on an M5-brane 
\smallskip
\eqn\mtwomfive{\matrix{\ \ &0&1&2&3&4&5&6&7&8&9&10\cr
M2&\hbox{X}&\hbox{X}&\hbox{X}&&&&&&&\cr
M5&\hbox{X}&\hbox{X}&&\hbox{X}&\hbox{X}&\hbox{X}&\hbox{X}&&&}}
\medskip
\noindent
it is necessary to switch on the $X^3$,$X^4$,$X^5$,$X^6$ scalar fields \BasuED. Given that these fields  depend  only on the  worldvolume coordinate $\sigma^2$,  the BPS condition  is    \BaggerSK\     
\eqn\bhbps{{dX^A\over d\sigma^2}\Gamma^A\Gamma^2\epsilon -{1\over 6}\varepsilon^{BCDA}\Gamma^A[X^B,X^C,X^D]\Gamma^{3456}\epsilon=0}
where $A,B,C,D=3,4,5,6$   and we used $\varepsilon^{ABCD}\Gamma^D=-\Gamma^{ABC}\Gamma_{3456}$.  For this field configuration the energy density is given by 
\eqn\hdbh{{\cal H}={1\over 2}\hbox{Tr} (\partial_2 X^A,\partial_2X^A)
+{1\over 12}\hbox{Tr}([X^A,X^B,X^C],[X^A,X^B,X^C]).} 
Following \BaggerVI, we write the potential as 
\eqn\dspo{V(X) ={1\over2}\hbox{Tr}\left({\partial W\over \partial X^A},{\partial W\over \partial X^A}\right)}
where
\eqn\supot{W ={1\over24}\varepsilon^{ABCD}\hbox{Tr}(X^A,[X^B,X^C,X^D]).}
Thus
\eqn\hdbhz{\eqalign{{\cal H}&={1\over 2}\hbox{Tr} \left(\partial_2 X^A+{\partial W\over \partial X^A},\partial_2X^A+{\partial W\over \partial X^A}\right)-\hbox{Tr}\left({\partial_2 X^A},{\partial W\over \partial X^A}\right)\cr
&={1\over 2}\hbox{Tr} \left(\partial_2 X^A+{\partial W\over \partial X^A},\partial_2X^A+{\partial W\over \partial X^A}\right)+{{\cal Z}^1\over 2}}}
where ${\cal Z}^1$ is the density of $Z_{\mu}^{(pq)}$, the 1-form central charge. Thus, for this field configuration ${\cal H}\ge {{\cal Z}^1\over 2}$ and the bound is saturated when 
\eqn\bhbpsf{{dX^A\over d\sigma^2} -{1\over 6}\varepsilon^{BCDA}[X^B,X^C,X^D]=0.}
When the \bhbpsf\ is satisfied, it follows  from \bhbps\  that the field configuration is half-BPS and the preserved supersymmetries satisfy $\Gamma^{2}\epsilon= \Gamma^{3456}\epsilon$. This is the configuration proposed by Basu and Harvey as the  M2-brane worldvolume soliton describing the branes system \mtwomfive.  In this section we have verified that the central charge associated to this state is the the 1-form central charge,  i.e. the central charge associated to  the M2-M5 intersection.  

\bigbreak\bigskip\bigskip\centerline{{\bf Acknowledgements}}\nobreak
We are grateful to Jaume Gomis for suggesting the problem and for useful discussions.  We also would like to thank Joaquim Gomis and Shunji Matsuura for fruitful conversations.   
This research was supported by Perimeter Institute for Theoretical
Physics.  Research at Perimeter Institute is supported by the Government
of Canada through Industry Canada and by the Province of Ontario through
the Ministry of Research and Innovation. We   also acknowledge further  support by an NSERC Discovery Grant.

\appendix{A}{Notation}
We summarize here our notation.  The indices are 
\eqn\summary{\eqalign{\hbox{worldvolume coordinates}:&\quad\mu,\nu=0,1,2\cr
\hbox{spatial worldvolume coordinates}:&\quad i, j=1,2\cr
\hbox{transverse space coordinates}:&\quad I, J=3,\ldots 10\cr
\hbox{$Spin(1,10)$ spinorial indices}:&\quad \alpha, \beta=1,\ldots 32\cr
\hbox{$Spin(1,2)$ spinorial indices}:&\quad \hat{\alpha}, \hat{\beta}=1,2\cr
\hbox{$Spin(8)$ chiral spinorial indices}:&\quad p,q,\dot{p},\dot{q} =1,\dots 8\cr
\hbox{${\cal A}$ algebra indices}:&\quad a,b=1,\ldots, \hbox{dim}{\cal A}}}
The Dirac matrices $\Gamma$ are a representation of the 11-dimensional Clifford algebra, i.e. given $m,n=0,\ldots, 10$ it results
\eqn\clif{\{\Gamma^m,\Gamma^n\}=2\eta^{mn}}
and 
\eqn\ccond{C^T=-C\qquad \Gamma_m^T=-C\Gamma_mC^{-1}.}
We take $\Gamma_m$ to be real matrices and $C=\Gamma^0$. The 11-dimensional spinors are Majorana (real) spinors with definite chirality respect to  $\Gamma_{012}$. Thus, they have 16  independent real components.    
\appendix{B}{Supercurrent Conservation}
We now show that the supercurrent \jz\ is conserved.  An easy computation gives 
\eqn\djz{\eqalign{\partial_\mu J^\mu=&-\partial_\mu(D_\nu X_a^I)\Gamma^\nu\Gamma^I\Gamma^\mu \Psi^a-D_\nu X_a^I\Gamma^\nu\Gamma^I\Gamma^\mu \partial_\mu\Psi^a\cr & -{1\over 2} \partial_\mu X_a^IX_b^JX_c^K f^{abcd}\Gamma^{IJK}\Gamma^\mu\Psi_d\cr & -{1\over 6} X_a^IX_b^JX_c^K f^{abcd}\Gamma^{IJK}\Gamma^\mu\partial_\mu\Psi_d.}}
Using the fundamental identity \fundid\ the previous equation can be rewritten as
\eqn\djzc{\eqalign{\partial_\mu J^\mu=&-(D_\mu D_\nu X_a^I)\Gamma^\nu\Gamma^I\Gamma^\mu \Psi^a-D_\nu X_a^I\Gamma^\nu\Gamma^I\Gamma^\mu D_\mu\Psi^a\cr & -{1\over 2} D_\mu X_a^IX_b^JX_c^K f^{abcd}\Gamma^{IJK}\Gamma^\mu\Psi_d\cr & -{1\over 6} X_a^IX_b^JX_c^K f^{abcd}\Gamma^{IJK}\Gamma^\mu D_\mu\Psi_d.}}
Inserting  the equations of motion \eomf\ and using  the  identity 
\eqn\fierz{\bar{\Psi}_c\Gamma^{IJ}\Psi_b\Gamma^I\Psi_aX_d^Jf^{cdba}=-\bar{\Psi}_c\Gamma_{\mu}\Psi_b\Gamma^\mu\Gamma^J\Psi_aX_d^Jf^{cdba},}
the right hand side of the \djzc\ results to be equal to zero.

\appendix{C}{Supersymmetry Variation of $J^0$}
In this appendix we compute the supersymmetry variation of  $J^0$, the zeroth component of the supercurrent \jz. Considering the ansatz $\Psi=0$ we get  
\eqn\varansn{\eqalign{\delta_{\epsilon}J^{0}=&-D_\mu X_a^ID_\nu X^{aJ}\Gamma^\mu\Gamma^I\Gamma^0\Gamma^\nu\Gamma^J\epsilon+{1\over 6}D_\mu X_a^IX_b^JX_c^KX_d^Mf^{bcda}\Gamma^\mu\Gamma^I\Gamma^0\Gamma^{JKM}\epsilon\cr&-{1\over 6}D_\mu X_a^IX_b^JX_c^KX_d^Mf^{bcda}\Gamma^{JKM}\Gamma^0\Gamma^\mu\Gamma^I\epsilon\cr
&+{1\over 36}X_a^IX_b^JX_c^KX_e^LX_f^MX_g^Nf^{abcd}f^{efg}{}_d \Gamma^{IJK} \Gamma^0\Gamma^{LMN}\epsilon.}}
We note that the right hand side of \varansn\ contains one term with two covariant derivatives, two terms  with one covariant derivative and one term without covariant  derivatives. Let's look first at the term  with two covariant derivatives.  Using the identity
\eqn\gamo{\eqalign{-\Gamma^\mu\Gamma^I\Gamma^0\Gamma^\nu\Gamma^J=&\Gamma^0\Gamma^{\mu\nu}\Gamma^{IJ}+\Gamma^0\Gamma^{\mu\nu}\delta^{IJ}+\Gamma^0\eta^{\mu\nu}\Gamma^{IJ}\cr &+\Gamma^0\eta^{\mu\nu}\delta^{IJ}-2\eta^{\mu 0}\Gamma^{\nu}\delta^{IJ}-2\eta^{\mu0}\Gamma^{\nu}\Gamma^{IJ}}}
we have 
\eqn\fp{\eqalign{-D_\mu X_a^ID_\nu X^{aJ}\Gamma^\mu\Gamma^I\Gamma^0\Gamma^\nu\Gamma^J\epsilon=&(D_0 X_a^I D_0 X^{aI}+D_i X_a^I D_i X^{aI})\Gamma^0\epsilon+2 D_0 X_a^I D_i X^{aI} \Gamma^i\epsilon \cr &+D_i X_a^I D_j X^{aJ}\Gamma^0\Gamma^{ij}\Gamma^{IJ}\epsilon.}} 
The  two terms with one  covariant derivative can be rearranged using the identity 
\eqn\gamt{-\Gamma^{\mu}\Gamma^{0}\Gamma^{I}\Gamma^{JKM}-\Gamma^{0}\Gamma^{\mu}\Gamma^{JKM}\Gamma^{I}=2 \eta^{\mu i}\Gamma^0\Gamma^i\Gamma^{IJKM}-6\eta^{\mu 0} \delta^{I[J}\Gamma^{KM]}}
and the last of the equations of motion \eomf. Thus we get 
\eqn\sp{\eqalign{&+{1\over 6}D_\mu X_a^IX_b^JX_c^KX_d^Mf^{bcda}\Gamma^\mu\Gamma^I\Gamma^0\Gamma^{JKM}\epsilon-{1\over 6}D_\mu X_a^IX_b^JX_c^KX_d^Mf^{bcda}\Gamma^{JKM}\Gamma^0\Gamma^\mu\Gamma^I\epsilon =\cr 
&+{1\over 3}D_i X_a^I X_b^JX_c^KX_d^M f^{bcda}\Gamma^0\Gamma^{i}\Gamma^{IJKM}\epsilon+{1\over 2}\varepsilon_{ij}\tilde{F}^{ijcd}X_c^IX_d^J\Gamma^{IJ}\epsilon.}}
Using the fundamental identity \fundid\ one can show that 
\eqn\afourx{\tilde{A}_i{}^g{}_a  X_g^I X_b^JX_c^KX_d^M f^{bcda}\Gamma^{IJKM}=0}
thus the \sp\ can be rewritten as 
\eqn\spdue{\eqalign{&+{1\over 6}D_\mu X_a^IX_b^JX_c^KX_d^Mf^{bcda}\Gamma^\mu\Gamma^I\Gamma^0\Gamma^{JKM}\epsilon-{1\over 6}D_\mu X_a^IX_b^JX_c^KX_d^Mf^{bcda}\Gamma^{JKM}\Gamma^0\Gamma^\mu\Gamma^I\epsilon =\cr 
&+{1\over 12}\partial_i (X_a^I X_b^JX_c^KX_d^M f^{bcda}\Gamma^0\Gamma^{i}\Gamma^{IJKM}\epsilon)+{1\over 2}\varepsilon_{ij}\tilde{F}^{ijcd}X_c^IX_d^J\Gamma^{IJ}\epsilon.}}

The term without covariant  derivatives can be simplified using the expression
\eqn\gt{\Gamma^{IJK}\Gamma^{LMN}=\Gamma^{IJKLMN}+9\Gamma^{[IJ}{}_{[MN}\delta^{K]}_{L]}+18\Gamma^{[I}{}_{[N}\delta^{K}_{L}\delta^{J]}_{M]}+6\delta^{[I}_{[N}\delta^{J}_{M}\delta^{K]}_{L]}} 
and the property of the $f^{abcd}$ structure constants. We get
\eqn\tp{\eqalign{&{1\over 36}X_a^IX_b^JX_c^KX_e^LX_f^MX_g^Nf^{abcd}f^{efg}{}_d \Gamma^{IJK} \Gamma^0\Gamma^{LMN}\epsilon=\cr&{1\over 6}\Gamma^0\epsilon X_a^IX_b^JX_c^KX_e^IX_f^JX_g^Kf^{abcd}f^{efg}{}_d=2\Gamma^0\epsilon V}} where $V$ is the potential defined in  \potential.

Collecting all the pieces together we have
\eqn\ncvar{\eqalign{\delta_{\epsilon}J^{0}=(D_0 X_a^ID_0 X^{aI}+D_i X_a^I D_i X^{aI}+2V)\Gamma^0\epsilon +2D_0 X_a^I D_i X^{aI}\Gamma^i\epsilon +\cr -\partial_i( X_a^I  D_j X^{aJ}\varepsilon^{ij}\Gamma^{IJ}\epsilon)+{1\over 12} \partial_i( X_a^I X_b^JX_c^KX_d^M f^{bcda}\Gamma^0\Gamma^{i}\Gamma^{IJKM}\epsilon)}} 
Considering the ansatz $\Psi=0$, the components of the stress-energy tensor \tmunu\  are  
\eqn\tensorsym{\eqalign{T_{00}=&{1\over 2}D_0 X_a^ID_0 X^{aI}+{1\over 2}D_i X_a^ID_i X^{aI}+V\cr T_{0i}=&D_0 X_a^I D_iX^{aI}}} 

Using  the   \tensorsym\ and the identity  $\Gamma^0\Gamma^i=-\epsilon^{ij}\Gamma^j\Gamma_{012}$ the \ncvar\ can be rewritten as 
\eqn\cvardue{\eqalign{\delta_{\epsilon}J^{0}=-2T^{0}{}_{\mu}\Gamma^{\mu}\epsilon  -\partial_i( X_a^I D_j X^{aJ}\varepsilon^{ij}\Gamma^{IJ}\epsilon)-{1\over 12} \partial_i( X_a^I X_b^JX_c^KX_d^M f^{bcda}\varepsilon^{0i}{}_\mu\Gamma^{IJKM}\Gamma^{\mu}\epsilon).}}

\listrefs

\end

It is known that the worldvolume supersymmetry algebra of the M2-brane includes  central charges. These central extensions appear in the anticommutator of the supercharges and  they are a 0-form and a 1-form for the worldvolume coordinates.  

It is known that the worldvolume supersymmetry algebra of the M2-brane include central charges. These central extension are associated with 
where
$\epsilon$ is the susy transformation parameter with $[\epsilon]=-{1\over 2}$